\documentclass[preprint,12pt]{elsarticle}

\usepackage{amssymb}
\usepackage[dvipsnames]{xcolor}
\usepackage[normalem]{ulem}

\definecolor{pierre}{RGB}{58, 92, 10}

\journal{Internet of Things}

\begin{document}

\begin{frontmatter}



\title{Low cost cloud based remote microscopy for biological sciences}


\author[inst1]{Pierre V Baudin\textsuperscript{*}\corref{.}}
\cortext[*]{These authors contributed equally to this work}
\affiliation[inst1]{organization={Department of Electrical and Computer Engineering, University of California Santa Cruz}}

\author[inst1]{Victoria T Ly\textsuperscript{*}\corref{.}}
\author[inst1]{Pattawong Pansodtee}
\author[inst1]{Erik A Jung}
\author[inst2]{Robert Currie}
\author[instRy]{Ryan Hoffman}
\author[inst4]{Helen Rankin Willsey}
\author[inst5,inst7]{Alex A Pollen}
\author[inst8,inst7]{Tomasz J Nowakowski}
\author[inst2,inst3,inst6]{David Haussler}
\author[inst2,inst5,inst7]{Mohammed Andres Mostajo-Radji}
\author[inst2,inst3,inst6]{Sofie Salama}
\author[inst1,inst2]{Mircea Teodorescu}

\affiliation[inst2]{organization={Genomics Institute, University of California, Santa Cruz}}
            
\affiliation[instRy]{organization={Department of Molecular, Cellular, and Developmental Biology, University of California, Santa Cruz}}

\affiliation[inst3]{organization={Department of Biomolecular Engineering, University of California, Santa Cruz}}
            
\affiliation[inst4]{organization={Department of Psychiatry and Behavioral Sciences, Weill Institute for Neurosciences, University of California, San Francisco}}
\affiliation[inst5]{organization={Department of Neurology, University of California, San Francisco}}

\affiliation[inst6]{organization=HHMI: Howard Hughes Medical Institute, University of California, Santa Cruz}

\affiliation[inst7]{organization=The Eli and Edythe Broad Center of Regeneration Medicine and Stem Cell Research, University of California, San Francisco}

\affiliation[inst8]{organization=Department of Anatomy, University of California, San Francisco}

\begin{abstract} 
A low cost remote imaging platform for biological applications was developed. The “Picroscope” is a device that allows the user to perform longitudinal imaging studies on multi-well cell culture plates. Here we present the network architecture and software used to facilitate communication between modules within the device as well as external cloud services. A web based console was created to control the device and view  experiment results. Post processing tools were developed to analyze captured data in the cloud. The result is a platform for controlling biological experiments from outside the lab.

\end{abstract}

\begin{graphicalabstract}
\includegraphics[width=\textwidth]{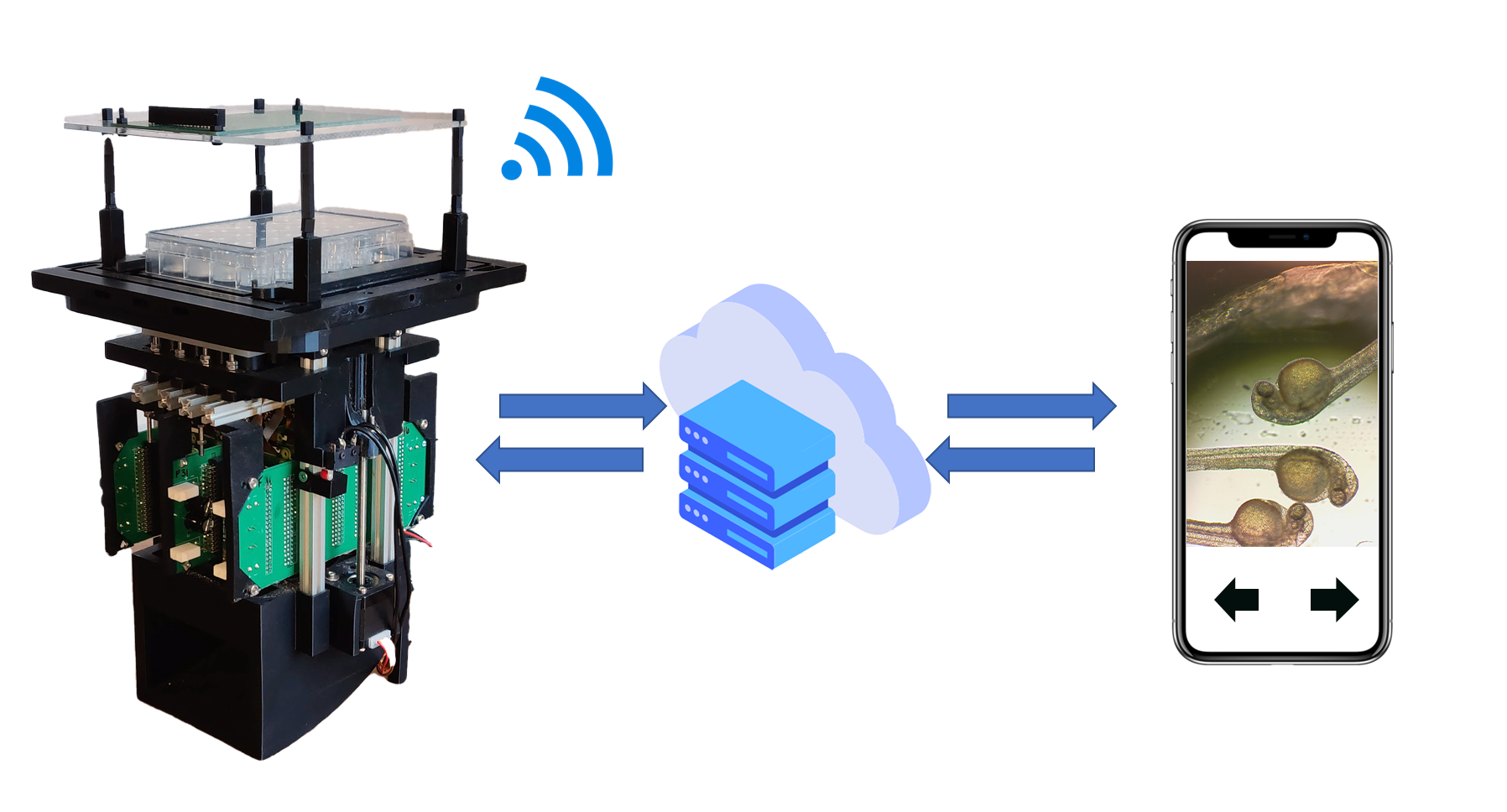}

\end{graphicalabstract}

\begin{highlights}
\item Multi-well parallel remote microscopy using primarily low cost off-the-shelf hardware.
\end{highlights}


\end{frontmatter}


\section{Introduction}
\label{sec:intro}
The COVID-19 pandemic has changed the work landscape throughout the world. Wherever possible jobs have transitioned to a remote format in compliance with lockdown regulations. Bench scientists were unequally affected by this pandemic,  experiencing a substantial reduction in 
their ability to work compared to computational scientists \cite{myers2020unequal}. This situation will likely have long-lasting effects on science careers, particularly for junior investigators. A silver lining may be the development of new approaches that allow experimental scientists to work remotely. These are likely to have lasting benefit long after the pandemic. We describe one such approach here.  

In addition to allowing a greater quantity of work to be done, remote experimentation and the automation required to implement it can increase the quality of work coming out of a lab. There is a ``crisis of reproducibility" that exists within biology \cite{sayre2018reproducibility, baker_biotech_2016}. Scientists and technicians following prescribed protocols are often unable to replicate each other's results. 
For example, in cell culture experiments, differences in how often controlled temperature and controlled gas incubators are opened or how much time a sample spends out of the incubator for routine manipulation can cause varying amounts of stress on the culture, affecting the metabolism and the experimental results. This leads to unaccounted for experimental variability. Increasing automation in lab experiments has been proposed as a way to address this issue. \cite{miles_achieving_2018}. Remotely operated experimentation entails such increased automation.

Techniques for remote operation exist on a wide spectrum of cost and complexity, from fully automated labs utilizing expensive robotic systems \cite{lippi_advantages_2019, evans_robotic_2008}, to DIY 3D printed microscopes with basic Internet access \cite{aidukas_low-cost_2019, maia_chagas_100_2017}. The further development of low cost solutions for remote lab control will bring more options within the reach of institutions with limited resources \cite{baden2015open}, allowing even labs in underprivileged environments to enjoy many of the benefits of the “lab of the future”\cite{miles_achieving_2018}. Many cost reductions have been made possible by the numerous innovations in the Internet of Things (IoT) space\cite{ornes2016core}, ranging from frameworks to low cost network capable devices\cite{chaczko_learning_2017, noauthor_node-red_nodate, lima2019performance, patel2019survey}.

There is an active community of DIY enthusiasts creating designs to manufacture lab equipment using consumer accessible tools \cite{may2019diy, meyer2012build,baden2015open}. Several open source microscope designs have been proposed using 3D printing and low cost computing platforms like the Raspberry Pi \cite{aidukas_low-cost_2019,maia_chagas_100_2017,10.5555/2535418.2535441}. Starting as low as 5\$ for a fully featured computer capable of running a desktop version of Linux, the Raspberry Pi allows scientists to use dedicated clusters of computers in virtually any application  \cite{abrahamsson2013affordable, saffran2016low, cox2014iridis}. 

Beyond research applications, remote and simulated lab systems have found use in educational environments \cite{gustavsson_visir_2007,alves_using_2011,blazquez-merino_use_2019}. Simulated labs have been used as a replacement for, or supplement to, traditional educational lab experience \cite{balamuralithara_virtual_2009,scheckler2003virtual} with the aim of introducing students without access to the necessary experimental equipment and environment to the experience of the scientific process. However, simulations can never provide students with the experience of actually discovering something new. Remote lab experimentation allows students to manipulate
live experiments running on real lab equipment from their classroom and home computers, or from their mobile phones. This removes the stale predictability of fully simulated experimentation, giving students a chance to experience the actual scientific process of discovery. Remote microscopy is an important aspect of many of these remote lab experiments \cite{jones2003learning,wallace_remote-access_2008, hossain2016interactive}. 

We recently described a device for simultaneous longitudinal imaging which we call the ``Picroscope" \cite{victoriapaper}.
Here we describe the software and network architecture developed to run the device as well as its integration into an IoT system on the cloud. This system enables adjustment of imaging parameters without disturbing the samples. It also allows researchers to monitor their experiment remotely enabling a variety of remote biology applications.

The Picroscope system is comprised of a cluster of network connected devices. A pipeline was developed to facilitate remote operation and to control communication between modules in the system. The result is a web based interface that allows users to control a longitudinal imaging experiment and view results in near real time. This brings high throughput parallel remote microscopy to a price point affordable in many sectors that could not previously access such systems. The 3D z-stack image data captured by the system allows it to image both 2D monolayer cell cultures and 3D samples. Our data pipeline is capable of feeding these z-stacks into software that generates Extended Depth of Field (EDoF) composite images \cite{http://bigwww.epfl.ch/publications/forster0401.html} to simplify the end user's visual analysis of longitudinal changes in a 3D sample. In this paper we demonstrate the system's functionality with frog embryos, zebrafish, and human cerebral cortex organoids.
\section{System Design}
\subsection{Overview}

\begin{figure}[!h]
    \includegraphics[width = 1\columnwidth]{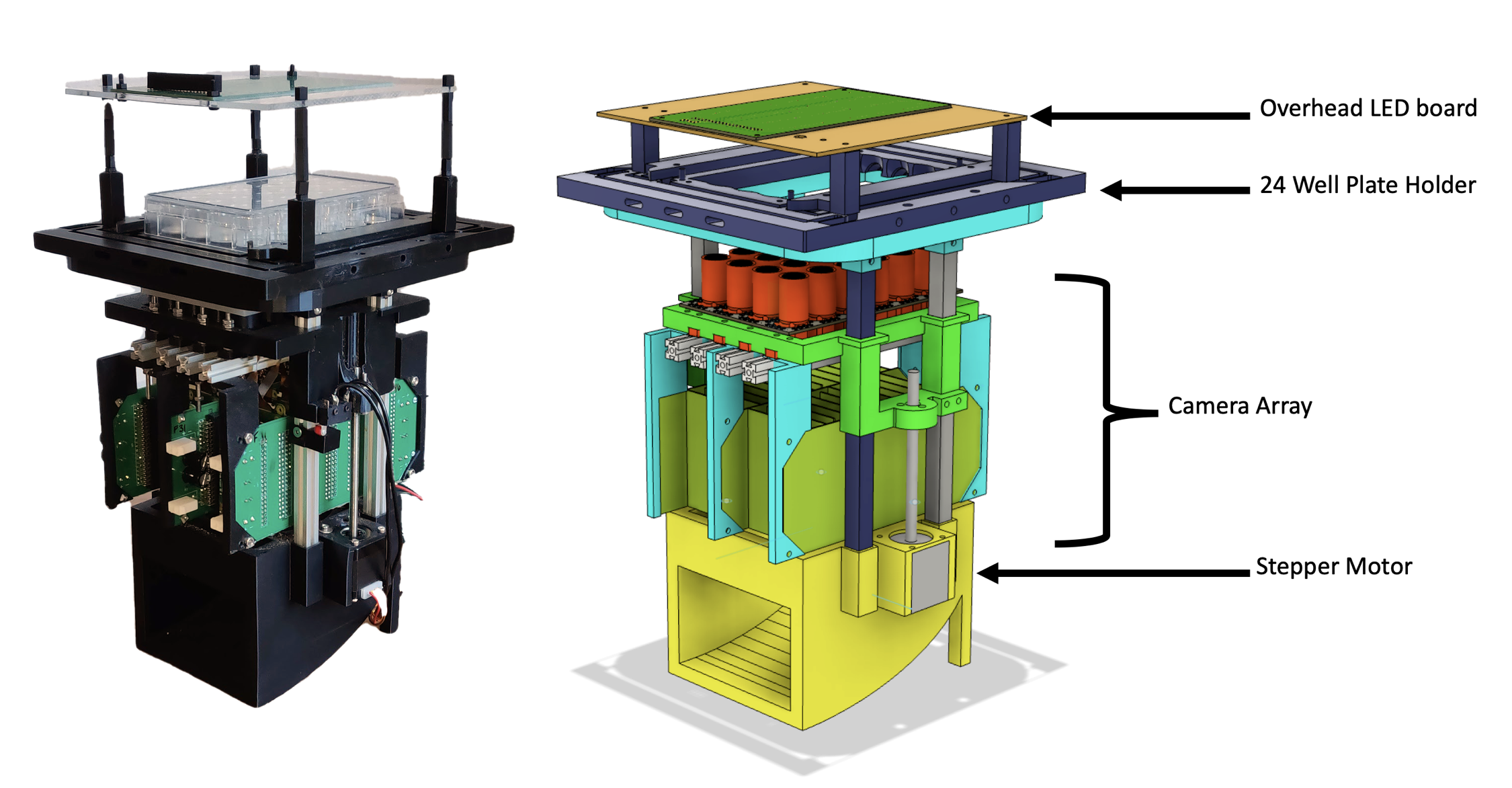}
    \caption{{\bf The Picroscope:} 
    The left shows a fully assembled Picroscope. On the right is the computer aided design(CAD) in Fusion 360 of the Picroscope (open source at (DH: give explicit link)). The main components of the picroscope include the over-the-plate illumination board, 24 well cell plate holder with XY stage, 24 camera array with stepper motor.}
   
    \label{fig:picroscope}
\end{figure}

The Picroscope contains several custom boards and 3D printed pieces. These are shown in figure \ref{fig:picroscope}, the main pieces include the 24 well plate holder, the elevator stage camera array, and the LED illumination boards (one above the culture plate, and one below). The camera array consists of a 6x4 grid of sensors with m12 threaded objective lenses attached. Two stepper motors are used to raise and lower the camera stage in order to move the focal plane. More detail on the hardware design of the picroscope can be found in \cite{victoriapaper}. 

\begin{figure}[!h]
    \includegraphics[width = 1\columnwidth]{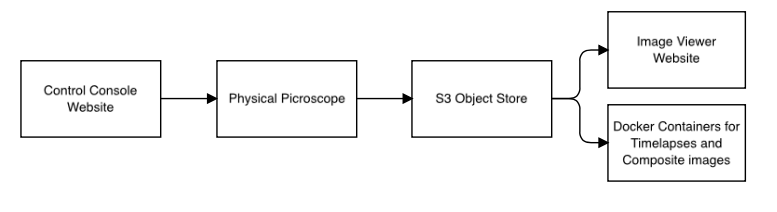}
    \caption{{\bf Basic workflow from user input to experiment output:}
    Control Console (Figure \ref{fig:console-and-viewer}), Physical Picroscope (Figure \ref{fig:physical-piscope-flowchart}), Image Viewer Website (Figure \ref{fig:console-and-viewer}), Docker Container (Figure \ref{fig:docker})
    \label{fig:top-levelflowchart}}
\end{figure}

\begin{figure}[!h]
    \includegraphics[width = 1\columnwidth]{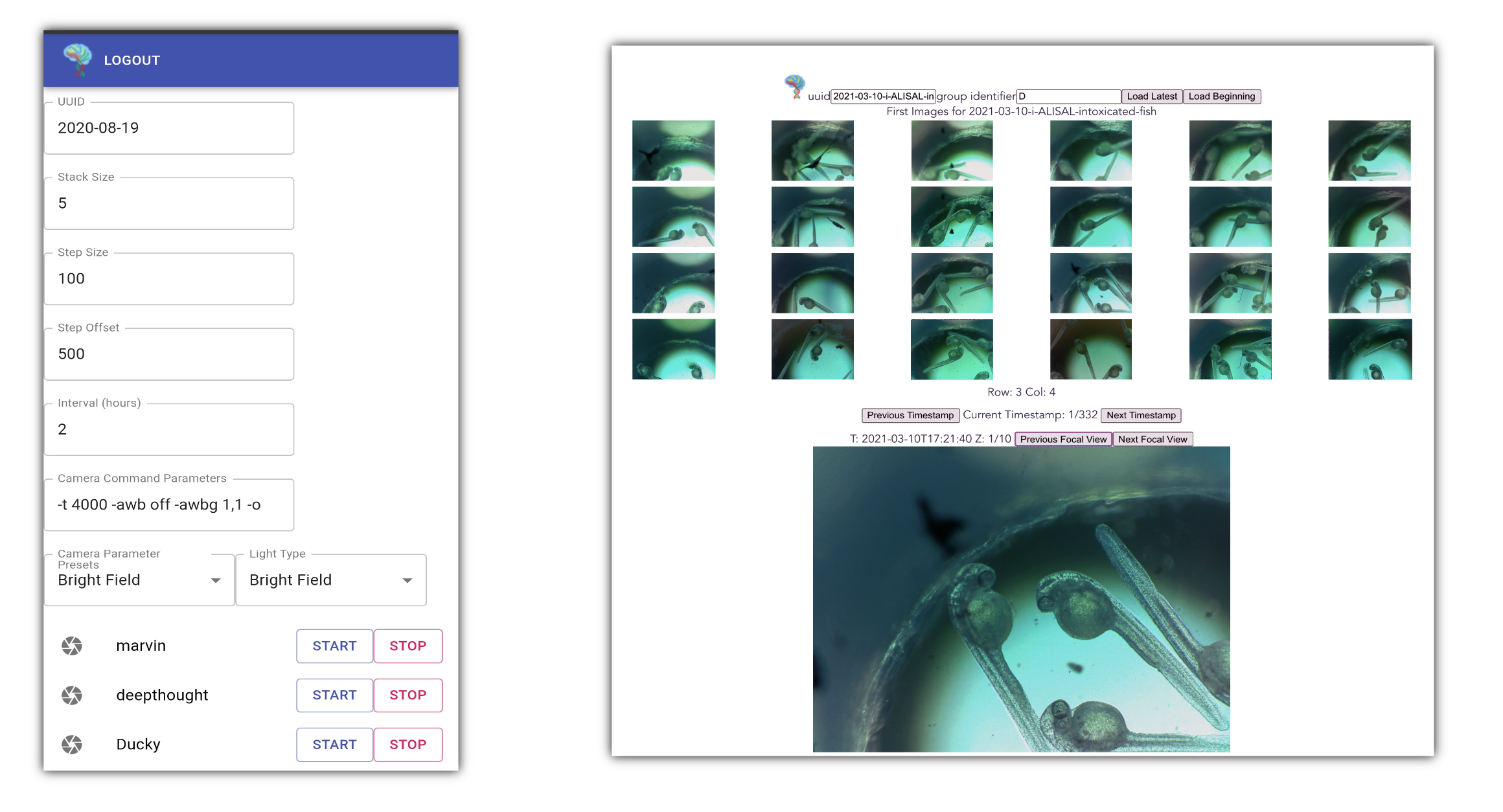}
    \caption{{\bf Web Interface:}
    Left: Control Console which allows us to set parameters and control the experiment, Right: Image Viewer where users can view their samples remotely }
    \label{fig:console-and-viewer}
\end{figure}

The basic workflow for this system is illustrated in figure \ref{fig:top-levelflowchart}). Experiments are triggered through our web based control console (Figure  \ref{fig:console-and-viewer}). In the console, the user sets the following parameters: experiment id, number of pictures in z-stack, distance between layers, initial offset distance, light type (Over-the-plate or Under-the-plate), and any additional camera control parameters allowed through the raspistill library \cite{noauthor_raspistill_nodate}. These parameters are passed to the Picroscope through a cloud based messaging service using the MQTT protocol\cite{locke2010mq}. Experiment parameters can also be changed on the fly during the course of an experiment through the same console.

An imaging event captured in our system consists of one z-stack per active camera. At the conclusion of each captured event, the Picroscope uploads the results to an S3 Object Store\cite{amazonS3} on a server where the pictures become accessible through our image viewer website (Figure \ref{fig:console-and-viewer}). Even though all 24 cameras share a single z-stack adjustment to reduce the cost of the device, the image viewer interface functionally provides users with 24 independent virtual microscopes. This allows users individual control with near real time views of the contents of each well at different z-stack focal layers. This is a big cost savings over having a classroom setup with 24 independent microscopes, for example. Students have the experience of controlling their own independent microscope from their mobile phone.  This illustrates a unique advantage of computer interfaced remote experimentation. 

\subsubsection{Device Hardware}

\begin{figure}[!h]
    \includegraphics[width = 1\columnwidth]{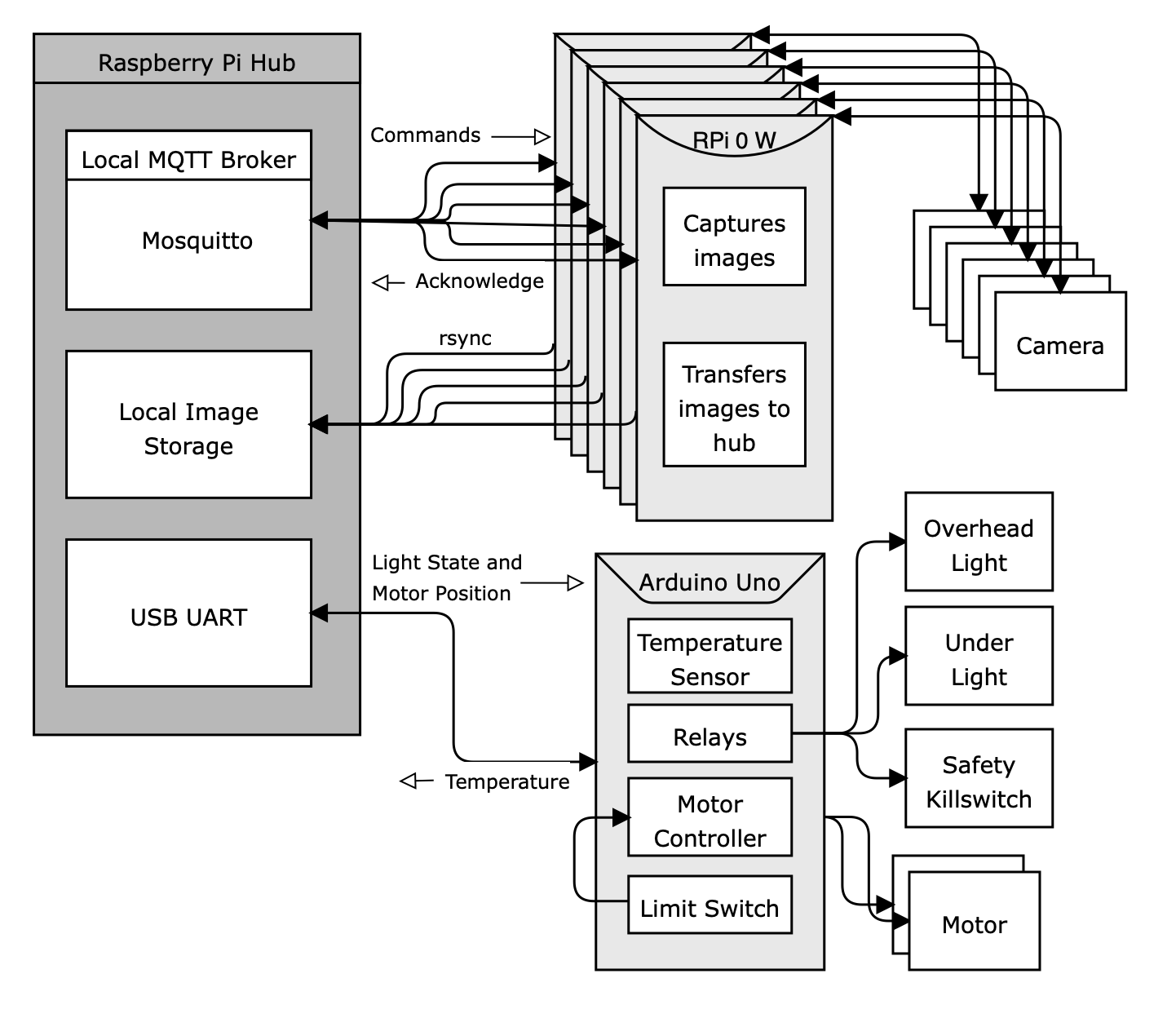}
    \caption{{\bf Data flow between local hardware network: }To gather a data point for an experiment, the hub sends commands to 24 Raspberry Pi Zero Ws each of which control one camera. The hub also connects to an Arduino Uno that is responsible for controlling motors and lights. In order to take a picture, the hub pi turns on the light and sends a command containing the input parameters for the raspistill camera API\cite{noauthor_raspistill_nodate} to the pi zeros. During a z-stack capture, pictures are taken and stored on the individual pi zeros. Each raspberry pi zero sends a message back to the hub when its picture has been taken. When all cameras have finished taking pictures, the camera stage is moved upwards by the motors, at which point the next layer of the z stack begins. At the conclusion of the z-stack capture, the stage lowers back to its starting position and the hub sends a command to each camera to begin transferring the images to the hub.}
    \label{fig:physical-piscope-flowchart}
\end{figure}

In addition to the custom pieces previously described, the Picroscope device is comprised of a number of connected sub-systems (Figure \ref{fig:physical-piscope-flowchart}) . Top level control is handled through the raspberry pi 4 based ``hub" which communicates with 24 raspberry pi zeros and an arduino uno. Figure \ref{fig:physical-piscope-flowchart} illustrates the communication between these devices.

\subsection{Communication Between System Layers}
\begin{figure}[!h]
    \includegraphics[width = 1\columnwidth]{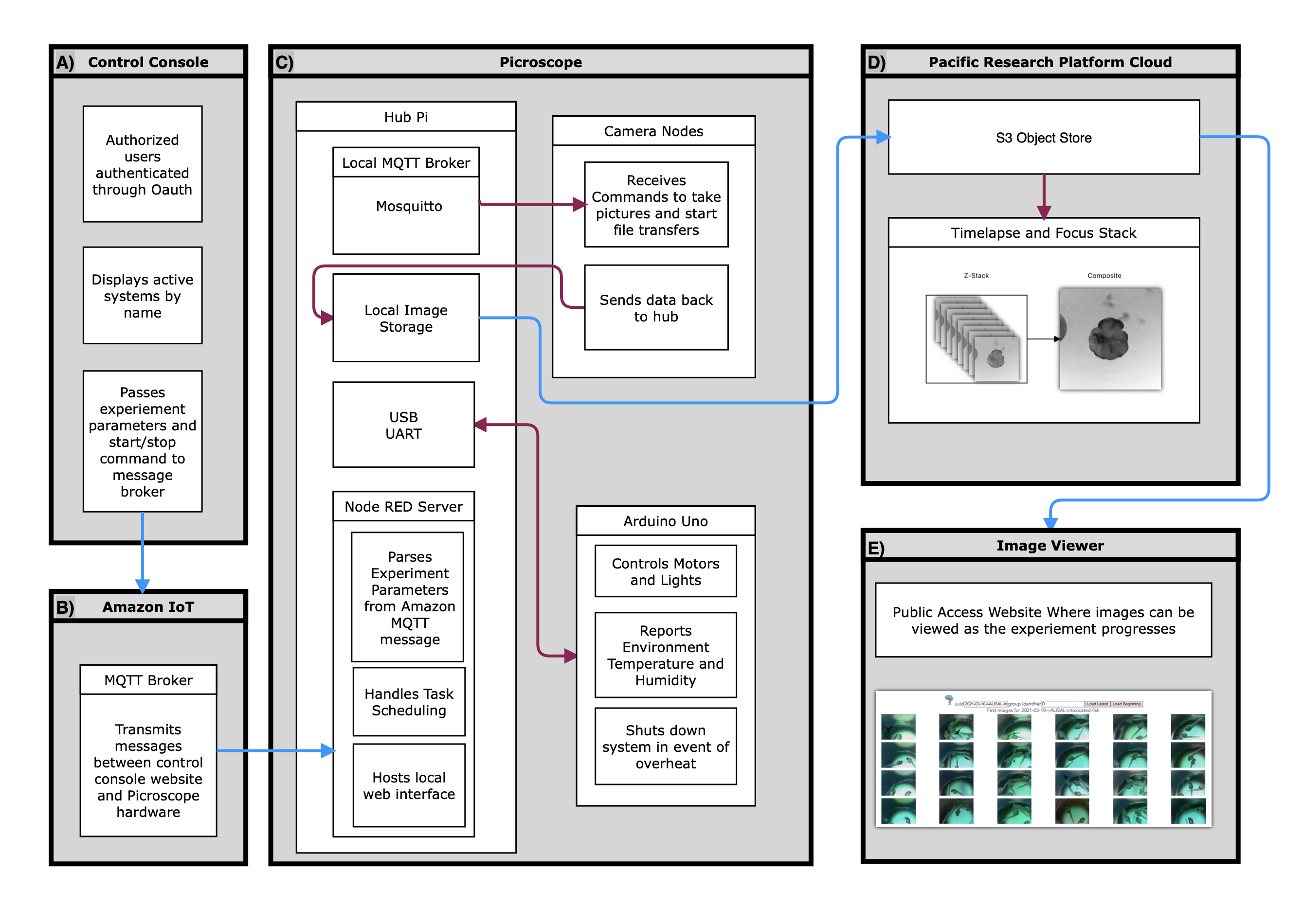}
    \caption{{\bf Message and data flow through the system:}
    Red lines represent communication within a local unit, Blue lines represent communication between physically separate subsystems. A)Control Console, B)Amazon IoT, C)Physical Picroscope, D)Cloud, E)Image Viewer}
    \label{fig:full-flowchart}
\end{figure}

The flow of messages and data in our pipeline is represented in Figure \ref{fig:full-flowchart}.

\noindent The control console webpage (figure \ref{fig:console-and-viewer}) communicates with our system using the MQTT message protocol. MQTT is a publish/subscribe based protocol in which a message “broker” transfers any messages published on a given topic to all the subscribers of that topic. To pass messages from the webpage to the picroscope, we use a cloud based MQTT broker provided by Amazon IoT. Every picroscope has a unique id and “device shadow” on the Amazon IoT platform. The control console website has access to the device list through Amazon’s IoT API. The console displays a list of all active systems with buttons to control each one (Figure \ref{fig:console-and-viewer}A). When a command is sent from the console it’s published with the topic being the id of the picroscope we wish to control.\\

The targeted picroscope receives the start command along with the desired experiment parameters. Parameters can be adjusted on the fly from the control console website. Each picroscope then uses it’s own locally hosted MQTT broker as a message bus to pass commands to the 24 raspberry pi zero Ws that each control a camera. Each hub also connects through USB to an Arduino Uno which is responsible for controlling the motors and lights as well as temperature safety monitoring and emergency shutoff.  Commands to take a picture can be sent to individual cameras or all of them at once. When a camera finishes taking a picture, it sends a message back to the hub with its camera id, allowing the hub to know when all cameras have finished. Unused wells can be disabled by the hub, allowing a higher maximum throughput for the other enabled cameras. 

\begin{figure}[!h]
    \includegraphics[width = 0.5\columnwidth]{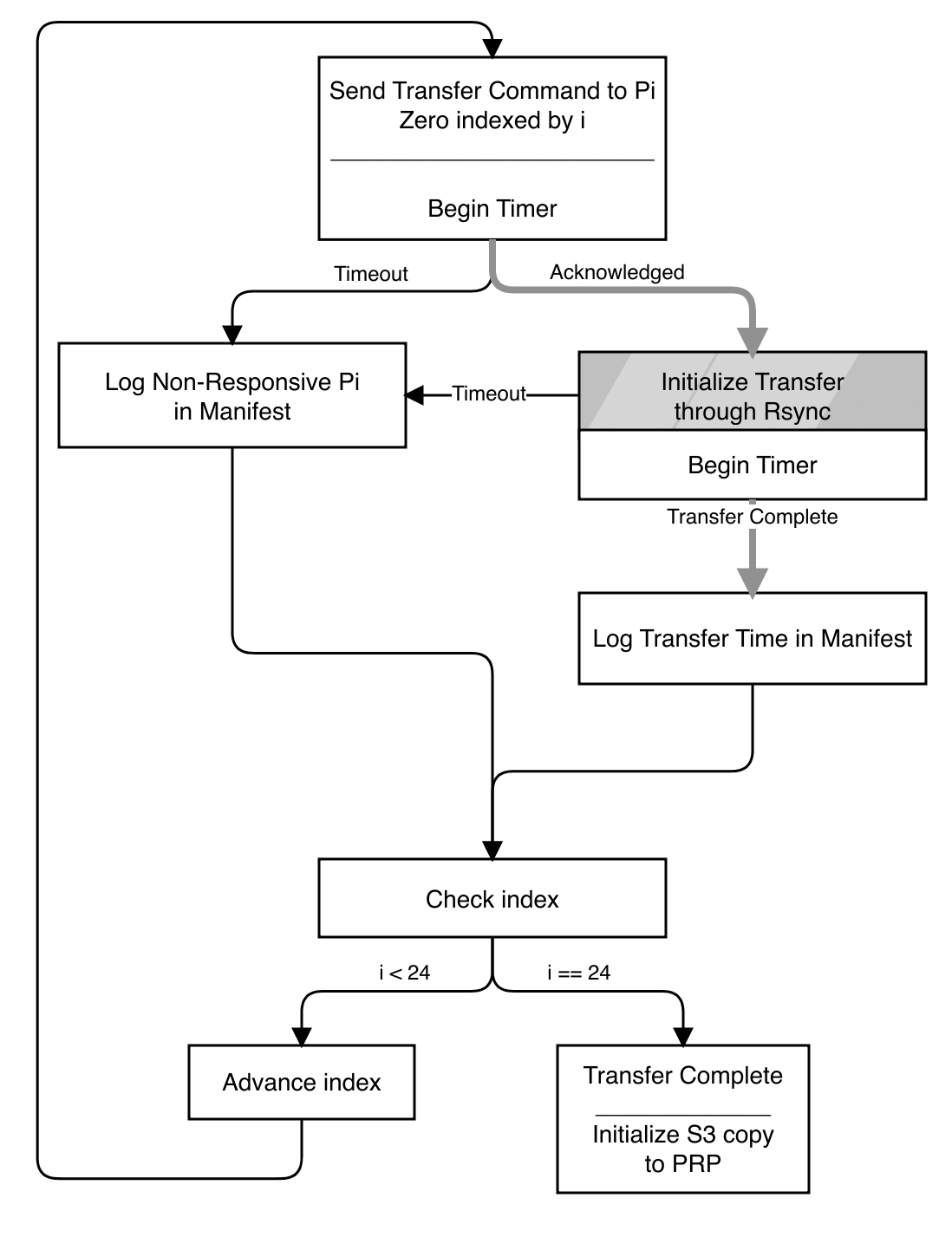}
    \caption{{\bf File Transfer Queuing Protocol:}
    White boxes with black arrows represent actions taken by the Pi Zeros, Grey Box and arrows indicate messages and actions taken by Pi Zero, Black arrows and Boxes indicate messages and actions taken by the Hub Pi}
    \label{fig:Transfer-Queue}
\end{figure}

When a z-stack capture concludes, the pi zeros need to send their data to their assigned hub pi. To accomplish this, we use a custom queuing protocol that initiates file transfers individually on each pi zero W and continues to the next pi when the current transfer finishes or a timeout condition is reached. The protocol is detailed in figure \ref{fig:Transfer-Queue}. This queuing system results in higher throughput than simply starting all transfers in parallel and the queue is not disrupted in the case of non-responsive pi zeros.
When the data transfer completes, the result is uploaded to an s3 object store on cloud hardware run on the Pacific Research Platform (or PRP) \cite{smarr2018pacific}. The time required for the transfer to complete is the primary limiting factor in determining the maximum data capture frequency we can achieve. Transfer time is primarily determined by z-stack size and number of active cameras. Using smaller z-stacks, or less cameras allows higher maximum throughput while maintaining parallel image capture for each well. This is an important consideration for imaging samples displaying higher frequency dynamics. With 24 cameras capturing 10 layer z-stacks, we are able to capture an entire new z-stack approximately 4 times per hour. If it is necessary to capture higher frequency dynamics, the picroscope can be set to capture short videos instead of z-stacks.


Once uploaded to s3, the images are accessible over the Internet. When an experiment is started, the picroscope generates a file called “manifest.json”. The manifest serves as a text based map indicating where in the s3 object store each image is stored. The manifest is updated with a new timestamp every time a new z-stack is captured. The manifest can be interpreted in such a way that you can generate the URLs for every picture without needing to query the object store (querying the object store takes a substantial amount of time). The image viewer website interprets the manifest to generate an interactive display of the images from a given experiment id. The manifest is also used when pulling the data into our dockerized scripts which we use to generate timelapse videos, focus stacked composite images and perform other image analysis.

\subsubsection{Timelapse Processing and Extended Depth of Field}

In addition to being viewable through our web interface, the pictures on the server are can be fed into scripts for generating focal plane stacked composite images with FIJI \cite{schindelin2012fiji} using the Extended Depth of Field plugin \cite{http://bigwww.epfl.ch/publications/forster0401.html}. This plugin allows us to generate a single image containing the best focused features from each layer of the z-stack. We can also use a script to generate timelapse videos from experiments. Generating timelapse videos with focus stacked frames allows for easy visual analysis of longitudinal changes in 3 dimensional samples.

The containerization of these programs with Docker allows us to run them in an automated fashion and easily deploy them with cloud service providers. \cite{10.5555/2600239.2600241}

\begin{figure}[!h]
    \includegraphics[width = 1\columnwidth]{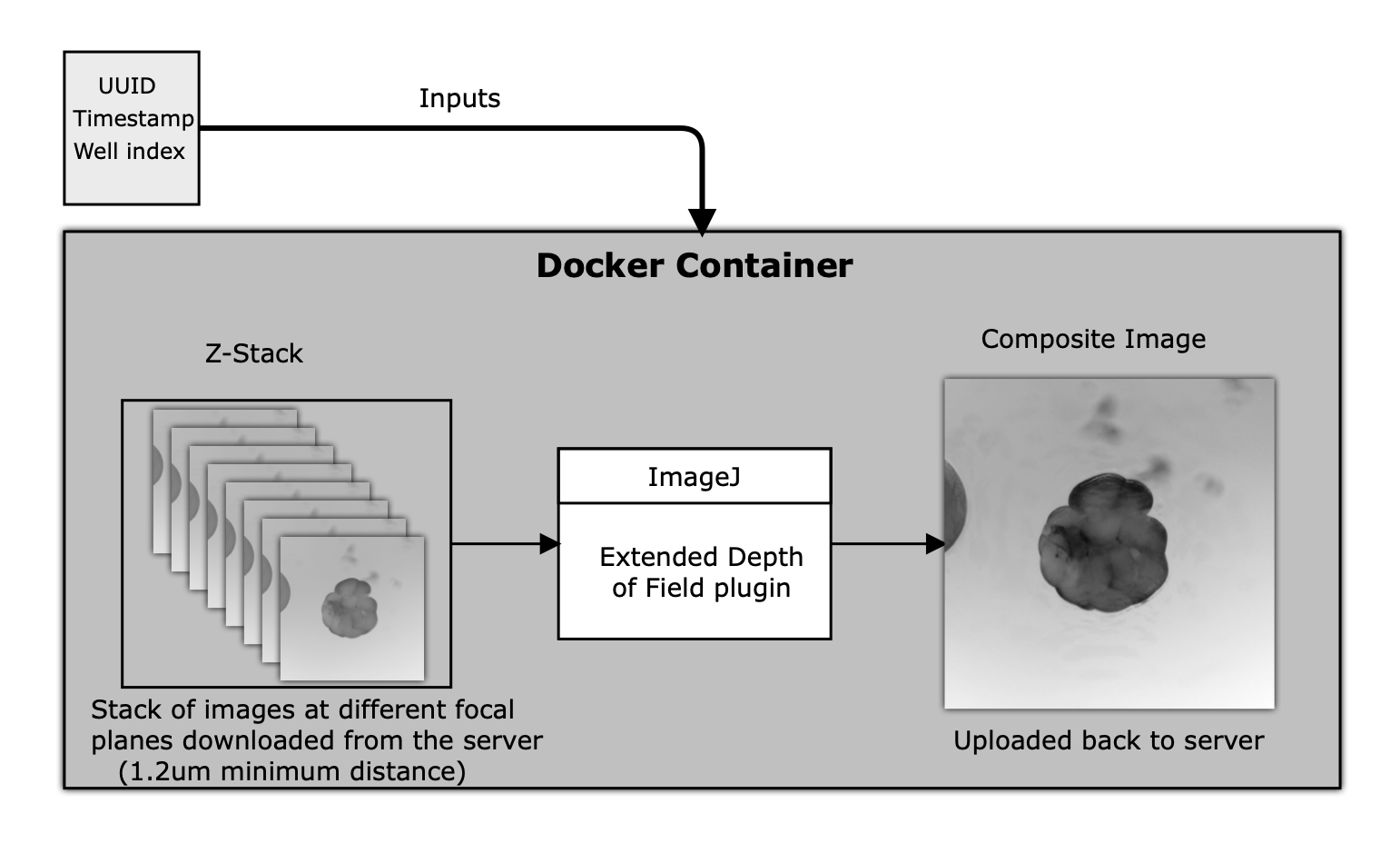}
    \caption{{\bf Extended Depth of Field:}
    Docker Containers are used to generate composite images with FIJI }
    \label{fig:docker}
\end{figure}

\section{Applications}
\subsection{In incubator application: visualizing 3D Human Brain Organoids}
\begin{figure}[!h]
    \includegraphics[width = 1\columnwidth]{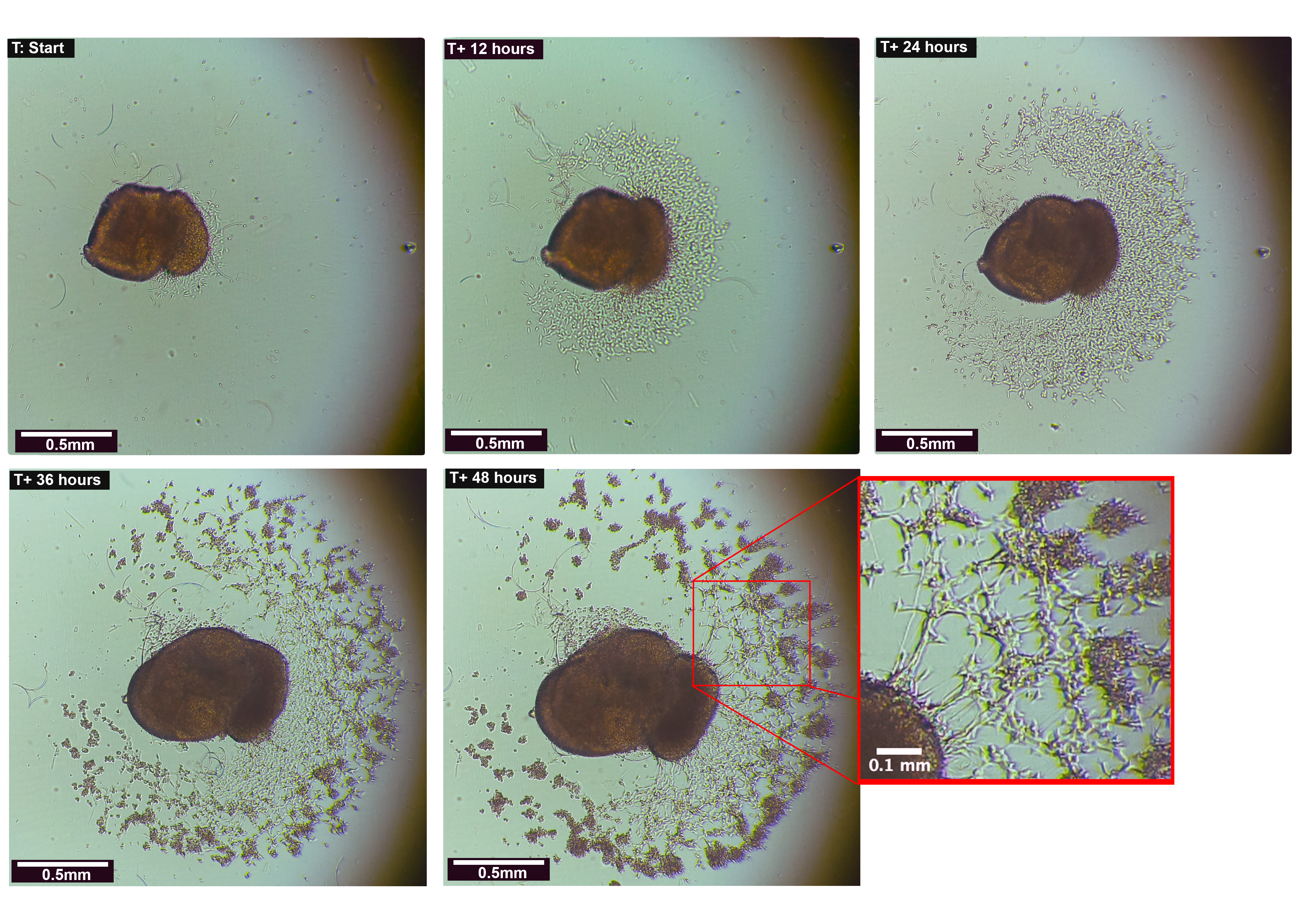}
    \caption{{\bf Human Cerebral Organoid with Outgrowths:}
    Organoids were plated on laminin coated 24 well plates and allowed to adhere to the surface. Outgrowths and cellular migration were restricted to the 2D plane of the plate’s surface and captured/monitored over the course of 21 days. Significant morphological changes occurred within the first 48 hours after imaging began.}
    \label{fig:neuron-results}
\end{figure}

Cerebral cortex organoids generated from aggregates of pluripotent stem cells have been shown to recapitulate aspects of early embryonic brain development, providing an in vitro model for studying species specific brain development in organisms including humans \cite{eiraku2008self,lancaster2014generation}. Here, human cerebral organoids were generated and plated on laminin coated 24 well plates and allowed to adhere to the surface. Outgrowths and cellular migration were restricted to the 2D plane of the plate’s surface and captured/monitored over the course of 21 days with the entire Picroscope device inside a temperature and humidity controlled CO2 incubator. Samples from that experiment can be seen in Figure \ref{fig:neuron-results}

\subsection{Live Whole Organism Imaging: Zebrafish}
In a small pilot program, 
we participated in a project to implement a simple experiment with live zebrafish to run remotely on the Picroscope
for a high school AP biology lab. The students used the Picroscope to measure survival and behavioral changes of zebrafish under the influence of varying concentrations of exogenous chemicals including caffeine and ammonia.

\begin{figure}[!h]
    \includegraphics[width = 0.75\columnwidth]{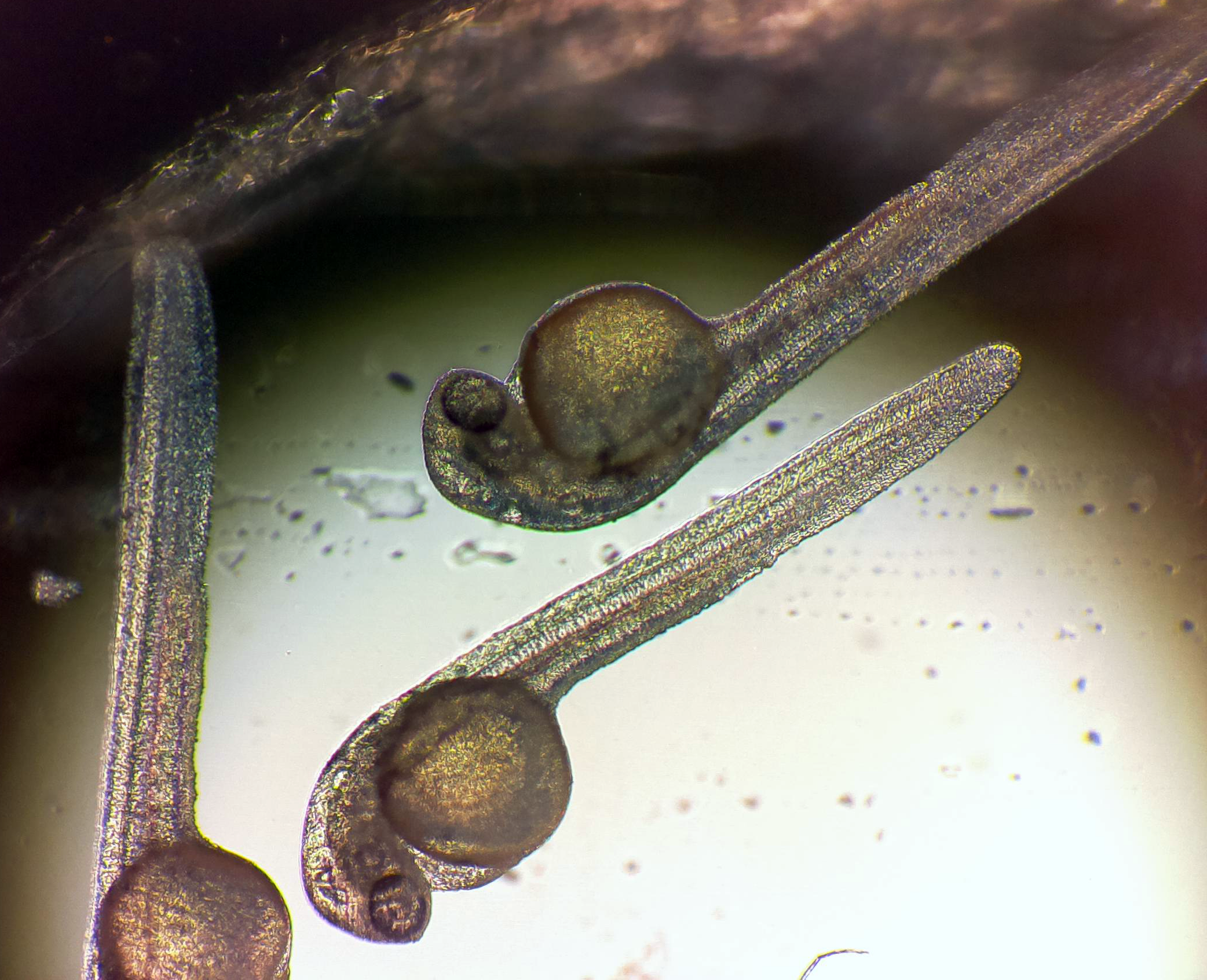}
    \caption{\bf{Zebrafish Circulation}
    Video included in supplemental material}

    \label{vid:zebrafish-video}
\end{figure}

In the set up phase of this experiment, we captured video showing fluid circulation inside a live zebrafish (figure \ref{vid:zebrafish-video}). This demonstrates the video capture capability which can be used to observe higher frequency dynamics than would be possible with z-stack image capture.

Feedback regarding the student experience was gathered using surveys with free response answers to questions. All users responded very positively to questions regarding usability, interest, and excitement in doing remote scientific experiments and indicated interest in pursuing future projects with the Picroscope. This experiment demonstrates the educational potential of remotely managing live whole organism studies on the Picroscope.

\subsection{Extended Depth of Field functionality: Frog Embryos}
\begin{figure}[!h]
    \includegraphics[width = 1\columnwidth]{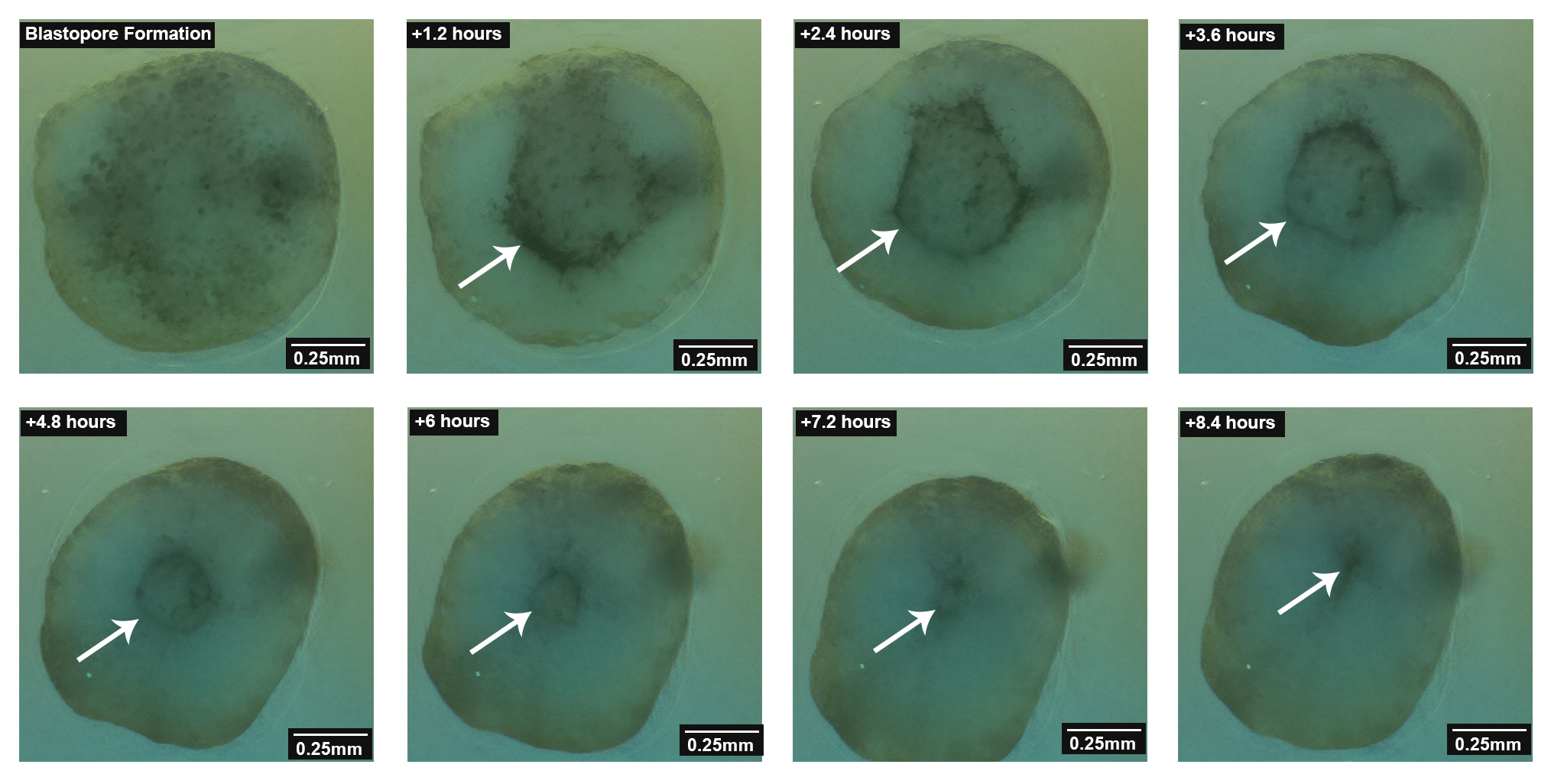}
    \caption{{\bf Blastopore Closure}
    Selected images from blastopore closure stage of development, these are samples from every 5 timesteps during this stage. the video in figure \ref{vid:composite-timelapse-D12} shows all frames.
    } 
    \label{fig:blast-closure}
\end{figure}

\begin{figure}[!h]
    \includegraphics[width = 1\columnwidth]{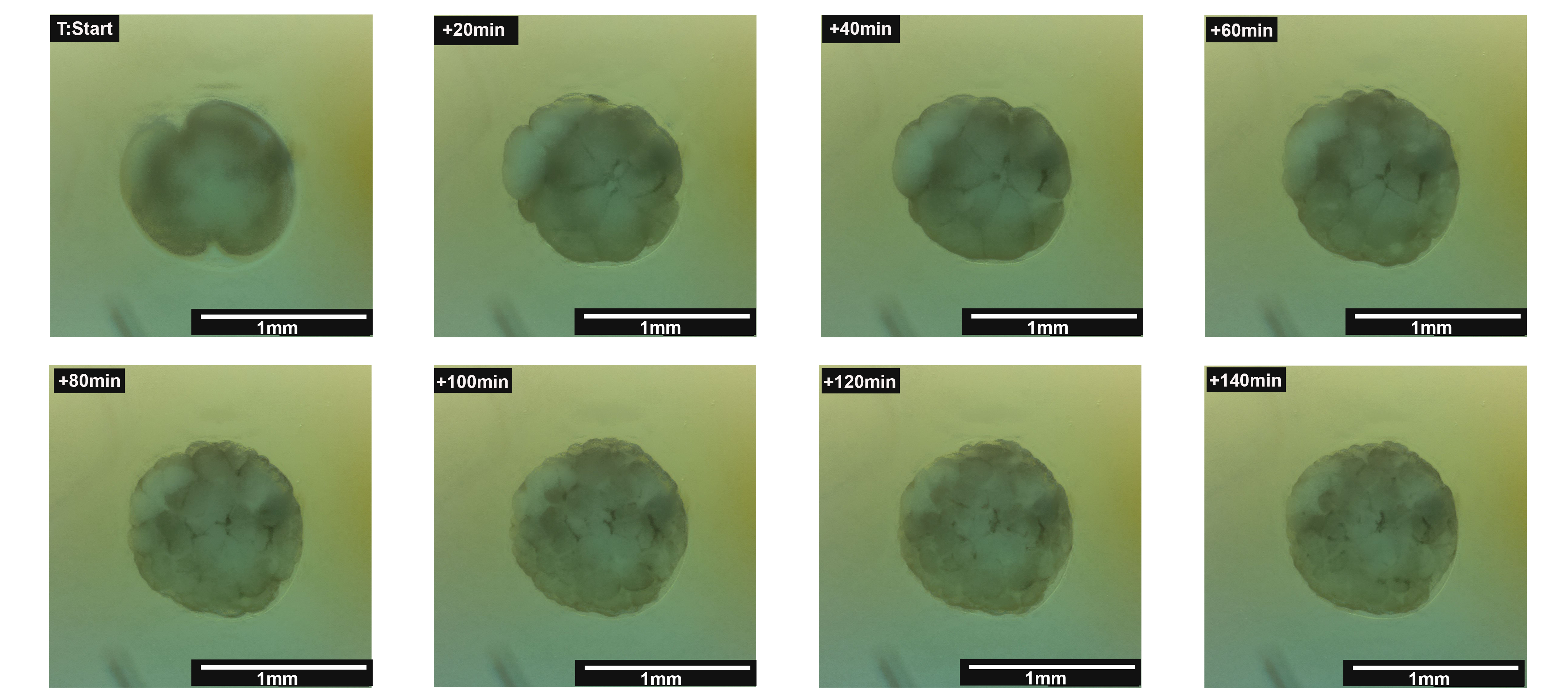}
    \caption{{\bf Early Cell Division}
    The first captured images in our data set show cell division as the embryo becomes more complex and individual cells shrink in size
    } 
    \label{fig:early-div}
\end{figure}

\begin{figure}[!h]
     \includegraphics[width = 0.5\columnwidth]{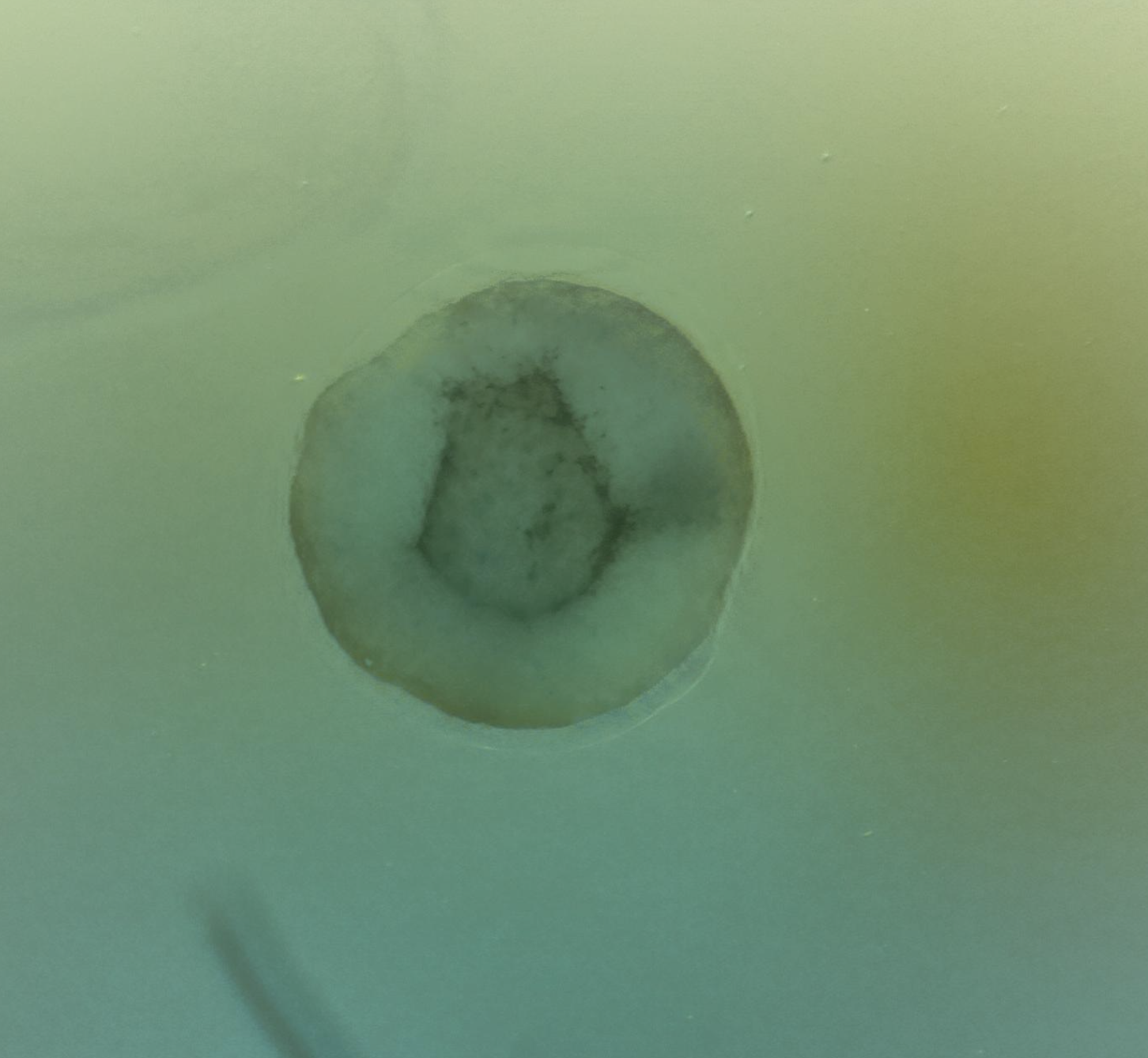}
    \caption{{\bf Frog embryo development timelapse video}
    Video included in supplemental material
    }
    \label{vid:composite-timelapse-D12}
\end{figure}

In another study, we monitored development of Xenopus Tropicalis frog embryos into larvae. We were able to image all stages of development as the embryo grew into a moving larva. Figures \ref{fig:blast-closure} and \ref{fig:early-div} show 2 interesting periods of development with different time scales of change. At the conclusion of this experiment, each z-stack timestep was fed through the Extended Depth of Field plugin in FIJI\cite{http://bigwww.epfl.ch/publications/forster0401.html}. Running FIJI through a docker container allowed the process to be scripted and run on a remote server. We then generated a timelapse video (figure \ref{vid:composite-timelapse-D12}) of these composite images. Each frame contains the in focus pieces of each of the 10 layers in the z-stack. When the embryo develops into the larval stage, it starts to move. The movement causes visible artifacts to appear in that section of the video, since the larvae moves between layer captures, demonstrating a drawback of this approach when imaging moving organisms.

\section{Discussion}
The Covid-19 pandemic changed the work landscape for many of us. The development of the picroscope was highly motivated by the access limitations we were presented with. The resulting system has helped our research group continue to produce work during this difficult time period.

The picroscope was developed as a modular device and can be deployed in a number of configurations optimized for different experimental settings. We have run experiments inside a standard CO2 incubator for up to 3 weeks at a time. We have demonstrated that our hardware is robust and minimally interferes with incubator environments.

The picroscope has been designed from the ground up as an extensible platform. Development of various compatible add-ons are in progress for new features including fluoresence microscopy. Our end goal is a general use parallel experiment system allowing remote control, sample manipulation, feeding, and imaging.

With this system we have provided a low cost solution (costing ~88\$ per well)\cite{victoriapaper} for biologists to work remotely with greater ease. We have developed a sensor-per-well parallel imaging system capable of brightfield microscopy that can be deployed inside a standard CO2 incubator. By having one camera per well, we have an array of microscopes available to researchers allowing them to remotely monitor the development of the biological samples over a long period of time.  

Having access to this system allows researchers to easily monitor long term morphological changes in their cell cultures without needing to interfere with their incubator environments. Using Picroscopes also allows for seamless collaboration between researchers at different institutions, allowing them to easily compare cultures as they grow. We envision deployment of many of these systems at once in our lab and collaborator's labs to help push us into an interconnected open source bio-lab of the future.

Anybody interested in building a picroscope system will find resources to do so at our research group's website http://braingeneers.gi.ucsc.edu/



 \bibliographystyle{elsarticle-num} 
 \bibliography{cas-refs}

\section*{ACKNOWLEDGMENTS}
This work is supported by the Schmidt Futures Foundation SF 857 (D.H.).
 Research reported in this publication was also supported by the National Institute Of Mental Health of the National Institutes of Health under Award Number R01MH120295 (S.R.S.) and the National Science Foundation under award number NSF 2034037 (M.T.).
We would like to thank Jeremy Linsley and Wiktoria Leks for providing us zebrafish for this study. 
In addition, we would like to thank Arnar Breevoort for providing experimental support. 
H.R.W. was supported by grant U01MH115747-01 from NIMH to Matthew State. M.A.M.-R. was partially supported by grant TL1 TR001871 from the NIH National Center for Advancing Translational Sciences. D.H. is an investigator with the Howard Hughes Medical Institute.

\section*{AUTHOR CONTRIBUTIONS}

P.V.B., V.T.L., P.P., worked on hardware and software of the Picroscope. R.C worked on the web console pipeline,  E.A.J. worked on early prototypes. V.T.L., P.V.B, H.R.W., R.H, M.A.M.-R. performed Biological experiments. H.R.W., R.H., M.A.M.-R., conceived the experiments.D.H., M.T., S.R.S, A.A.P. M.A.M.-R., M.R. and T.J.N. supervised the team and secured funding. P.V.B., V.T.L., M.A.M.-R., M.T. wrote the manuscript with contributions from all authors.

\section*{COMPETING INTERESTS}
The authors have written patents covering the technology described in this article. A.A.P. is on the board of Herophilus.  The authors declare no other conflict of interest.





\end{document}